\documentclass[a4paper,usenatbib]{mnras}
\usepackage{aas_macros}
\usepackage{ae,aecompl}
\usepackage{graphicx}
\usepackage{amsmath}
\usepackage{amssymb}
\usepackage[dvipsnames]{xcolor}
\hypersetup{colorlinks=true,allcolors=teal}
\usepackage{microtype}
\newcommand{\Msun}{\ensuremath{M_{\odot}}}
\newcommand{\Mh}{\ensuremath{h^{-1}M_{\odot}}}
\newcommand{\Mhsq}{\ensuremath{h^{-2}M_{\odot}}}
\newcommand{\Mpch}{\ensuremath{h^{-1}{\rm Mpc}}}
\newcommand{\avg}[1]{\ensuremath{\left< #1 \right>}}

\newcommand{\der}{\ensuremath{{\rm d}}}

\newcommand{\erf}[1]{\ensuremath{{\rm erf}\left(#1\right)}}

\newcommand{\mHI}{\ensuremath{m_{\textsc{Hi}}}}
\newcommand{\Hi}{\textsc{Hi}}
\newcommand{\eqn}[1]{equation~\eqref{#1}}

\newcommand{\be}{\begin{equation}}
\newcommand{\ee}{\end{equation}}
\newcommand{\ph}[1]{\phantom{#1}}

\title[Halo models of HI selected galaxies]{Halo models of HI selected galaxies}
\author[Paul, Choudhury \& Paranjape]{
Niladri Paul,$^{1}$\thanks{E-mail: npaul@iucaa.in}
Tirthankar Roy Choudhury$^{2}$\thanks{E-mail: tirth@ncra.tifr.res.in}
\& Aseem Paranjape$^{1}$\thanks{E-mail: aseem@iucaa.in}
\\
$^{1}$Inter-University Centre for Astronomy and Astrophysics, 
     Ganeshkhind, Post Bag 4, Pune 411007, India.\\
$^{2}$National Centre for Radio Astrophysics, Tata Institute of Fundamental Research, 
     Ganeshkhind, Post Bag 3, Pune 411007, India.
}
\date{draft}
\begin{document}
\label{firstpage}
\pagerange{\pageref{firstpage}--\pageref{lastpage}}
\maketitle
\begin{abstract}
%%%%%%%%%%%%%%%%%%%%%%%%%%%%%%%%
Modelling the distribution of neutral hydrogen (\textsc{Hi}) in dark matter halos is important for studying galaxy evolution in the cosmological context. %21 words
We use a novel approach to infer the \textsc{Hi}-dark matter connection at the massive end ($m_{\textsc{Hi}} > 10^{9.8} \Msun$) from radio \textsc{Hi} emission surveys, using optical properties of low-redshift galaxies as an intermediary. %33 words
In particular, we use a previously calibrated optical HOD describing the luminosity- and colour-dependent clustering of SDSS galaxies and describe the \textsc{Hi} content using a statistical scaling relation between the optical properties and \textsc{Hi} mass. %36 words
This allows us to compute the abundance and clustering properties of \textsc{Hi}-selected galaxies and compare with data from the ALFALFA survey. %22 words
We apply an MCMC-based statistical analysis to constrain the free parameters related to the scaling relation. %17 words
The resulting best-fit scaling relation identifies massive \textsc{Hi} galaxies primarily with optically faint blue centrals, consistent with expectations from galaxy formation models. % 23 words 
We compare the {\Hi}-stellar mass relation predicted by our model with independent observations from matched {\Hi}-optical galaxy samples, finding reasonable agreement. As a further application, we make some preliminary forecasts for future observations of HI and optical galaxies in the expected overlap volume of SKA and Euclid/LSST. % 50 words
%%%%%%%%%%%%%%%%%%%%%%%%%%%%%%%%
%% Total 202 words
%%%%%%%%%%%%%%%%%%%%%%%%%%%%%%%%
\end{abstract}
\section{Introduction}
\label{sec:intro}
\noindent
In the current understanding of Lambda-cold dark matter ($\Lambda$CDM) cosmology, the large-scale structure of the universe forms in a hierarchical manner driven by the collisionless cold dark matter. The gravitational instability leads to the formation of high density collapsed objects (or haloes) that are decoupled from the cosmic expansion. The baryons, being less abundant, follow the dark matter distribution and settle in the gravitational potential well formed by the dark matter haloes and eventually form stars and galaxies \citep{Mo_White_Bosch_book_2010}. Perhaps the most effective observational probes of structure formation, particularly at low redshifts (say $z \lesssim 0.1$), are related to the stars and the interstellar medium of these galaxies. For example, the environmental dependence of star formation in galaxies can be probed using optical and UV (ultra-violet) luminosities through large surveys like the 2dF Galaxy Redshift Survey (2dFGRS) \citep{2DFGRS_Norberg_et_al_2001,2dFGRS_Lewis_et_al_2002} and Sloan Digital Sky Survey (SDSS) \citep{Baldry_et_al_2004,Zehavi_et_al_2005,ss09}. Alternatively, one can also probe the low-density gas residing in the interstellar medium using atomic and molecular lines, one of the most prominent lines being the 21~cm transition of neutral hydrogen (\Hi) \citep{Draine_book_2011}. At low redshifts, this line is detected in emission through surveys like the HI Parkes All-Sky Survey (HIPASS) \citep{Zwaan_et_al_2003, Zwaan_et_al_2005_a} and Arecibo Legacy Fast ALFA (ALFALFA) \citep{ALFALFA_giovanelli_series_I, ALFALFA_giovanelli_series_II, ALFALFA_giovanelli_series_III, ALFALFA_giovanelli_series_IV} which can provide us with a robust estimate of the \Hi\ mass of the galaxy. The cold gas traced by the \Hi\ 21~cm line can act as a natural source for star formation and thus one expects some correlation between the \Hi\ and optical properties of the galaxies. Such correlations have been explored by cross-matching the \Hi\ galaxy catalogues with samples of optical and ultra-violet selected galaxies and studying physically interesting quantities such as the \Hi-to-stellar mass ratio \citep{Catinella_et_al_2010,maddox+15}. Recently \citet{Rafieferantsoa_et_al_2018} have used machine learning techniques to predict the \Hi\ content of galaxies based on their optical properties.  The environmental dependence of \Hi\ content has also been studied both observationally \citep{Jones_et_al_2016, Stark_et_al_2016} and theoretically \citep{Cunnama_et_al_2014, Rafi_et_al_2015}.
On the theoretical front, the formation and evolution of dark matter haloes are reasonably well understood, thanks to both $N$-body simulations and analytical calculations. Since these dark matter haloes cannot be probed directly via observations, one needs to include some basic understanding of the galaxy formation process for connecting the theoretical calculations to the observational data. Unfortunately, the theoretical modelling of galaxy formation consists of various complex astrophysical processes, many of which are relatively poorly understood. These include, e.g., gas cooling, fragmentation, formation of molecular gas, star formation, feedback processes etc. These processes are highly non-linear and cover a wide range of length and temporal scales. Hence it is almost impossible to model them self-consistently in analytical calculations or numerical simulations. The most common approach in modelling the galaxy formation has been to make approximations in the calculations by introducing a number of free parameters. These parameters can subsequently be constrained by matching with observations over a wide range of wavebands, thus gaining insights into the physical processes involved \citep[see][for a review]{Somerville_Dave_2015}.
An alternate, statistical approach in connecting the observations of galaxies to the dark matter haloes is to use a phenomenological prescription to populate the haloes with galaxies. These prescriptions consist of free parameters which are constrained using observations. These include, e.g., the Halo Occupation Distribution   \citep[HOD; see, e.g.,][for a review]{Cooray_Sheth_2002} and SubHalo Abundance Matching (SHAM) based approaches \citep[see, e.g.,][]{Reddick_et_al_2013,hearin+13}, both of which have been widely used while studying the optical/UV-selected galaxies. The main advantage of these methods is that they do not require any assumptions about the astrophysical processes (say, cooling or feedback) inside the halo, rather the physical insights can be inferred indirectly once the model parameters are reliably constrained \citep{kravtsov+04,zheng+05}.
Since the number of \Hi-selected galaxies has become quite substantial in recent years, it is worth exploring whether such statistical models can be naturally extended to understand the \Hi\ properties of galaxies \citep{Wyithe_et_al_2009, Wyithe_Brown_2010, Padmanabhan_et_al_2016, Padmanabhan_refregier_2017, Padmanabhan_et_al_2017, Padmanabhan_kulkarni_2017}. Studies along these lines have been attempted recently by  \citet{Guo_et_al_2017} who have found that a SHAM-based model can match both the mass function and two-point correlation of the \Hi\ galaxies.
In this paper, we perform an independent investigation of HOD-based models of the \Hi\ distribution, using the abundance and clustering data of resolved \Hi\ galaxies at low redshifts. Our analysis, based on the Markov chain Monte Carlo (MCMC) method is slightly different from standard parametric approach of HOD \citep{Zehavi_et_al_2005, Zehavi_et_al_2011, Guo_et_al_2015}. We have tried the parametric HOD approach which gives unphysical results as already pointed out by \citet{Guo_et_al_2017}.  So we use a novel treatment where we combine pre-calibrated optical HODs -- these describe the luminosity- and colour-dependent clustering of SDSS galaxies \citep{Zehavi_et_al_2011,ss09, Guo_et_al_2015} -- with a modelled scaling relation between the optical properties and \Hi\ mass of the galaxies. The free parameters of the scaling relation can be constrained using observations of \Hi\ galaxies, thus allowing us to infer the correlation between the optical properties and \Hi\ content. Note that this analysis does \emph{not} require the \Hi\ and optical samples to be necessarily matched and thus can contain a larger number of objects. One can, of course, use the results of the matched samples as additional constraints in our analysis which may further reduce the uncertainties in the values of the free parameters.
    
The article is organised as follows. We describe the data set used in this work in section~\ref{sec:data}. In section \ref{sec:scaling}, we present our analysis based on scaling relation between the \Hi\ mass and optical properties of the galaxies.  We present a detailed discussion of our results in section \ref{sec:discuss} which also includes the prediction for upcoming surveys like SKA, LSST etc. in subsection \ref{sec:prediction}. , and conclude in section \ref{sec:conclude}. Some of the details of our calculations are given in the Appendices.
Throughout this article, we have chosen a flat $\Lambda$CDM cosmology with matter density parameter $\Omega_m = 0.307$, baryon density parameter $\Omega_b = 0.048$, Hubble parameter $H_0 = 100~h~\rm km~s^{-1} Mpc^{-1}$ with $h = 0.678$, primordial r.m.s. density fluctuations at a scale of $8 \Mpch$, $\sigma_8 = 0.823$ and an inflationary spectral index, $n_s = 0.96$, consistent with the results of Planck \citep{Planck_2014}, and also with the values assumed by \citet{Guo_et_al_2015} in calibrating their optical HOD, which we will use in our scaling analysis. Wherever needed, we have used the \citet{Eisenstein_Hu_1999} fitting function for the linear theory matter power spectrum. We will quote halo masses in $\Mh$, neutral hydrogen masses in $\Msun$ and stellar masses in $\Mhsq$. Galaxy luminosities are quoted in terms of $M_r \equiv M_{0.1r} - 5 \log_{10}(h)$, where $M_{0.1r}$ is SDSS $r$-band absolute magnitude, K-corrected and evolution corrected to $z = 0.1$ \citep{Blanton_et_al_2003}. 
\section{Data set}
\label{sec:data}
\noindent
In this work, we use the abundance and clustering data of individually resolved galaxies based on their \Hi\ mass from ALFALFA survey \citep{Guo_et_al_2017} to constrain our scaling-based halo model. \footnote{The correlation measurements and their covariance matrices can be downloaded from the website of Hong Guo \href{http://sdss4.shao.ac.cn/guoh/}{http://sdss4.shao.ac.cn/guoh/}} The sample consists of $\approx 16000$ \Hi-selected galaxies from the $\alpha.70$ \footnote{\href{http://egg.astro.cornell.edu/alfalfa/data/index.php}{http://egg.astro.cornell.edu/alfalfa/data/index.php}}  catalogue of the ALFALFA survey \citep{ALFALFA_giovanelli_series_I}.  The sample covers a redshift range of $0.0025<z<0.05$ with an area of 4693 $\rm deg^2$ in sky resulting in a total comoving volume of $1.55 \times 10^6 h^{-3}\rm Mpc^{3}$.  In particular, \citet{Guo_et_al_2017} provides the abundance and clustering data of the \Hi\ sample for a total of 13 \Hi\ mass thresholds starting from $\log [\mHI/\Msun]>8.0$ to $\log [\mHI/\Msun]>10.4$. Due to incompleteness of the SDSS sample which we discuss in section \ref{sec:sample_completeness}, we only use the \Hi\ abundance and clustering information for $\log [\mHI/\Msun]>9.8$. 
For the scaling relation analysis, we also use the optical HOD prescribed in \citet[][see their Table 2]{Guo_et_al_2015}. This optical HOD was calibrated using the optical galaxy sample of the New York University Value-Added Galaxy Catalog (NYU-VAGC; \citep{Blanton_et_al_2005}) which in turn was constructed from the SDSS Data Release 7\footnote{\href{http://classic.sdss.org/dr7/}{http://classic.sdss.org/dr7/}} Main galaxy sample \citep{Abajajian_et_al_2009}. The DR7 sample covers an effective sky-area of about $7300 \rm deg^2$ with redshift range $0.02<z<0.25$ covering a comoving volume $\sim 0.26 h^{-3}\rm Gpc^{3}$. The optical HOD calibrations are available only for magnitudes $M_r < -18.0$.  
\section{Scaling between HI and optical properties}
\label{sec:scaling}
\noindent
In this section, we discuss a novel approach to understand the relation between \Hi\ content and the halo mass, by exploring the relation between \Hi\ and optical properties of the galaxies. To this end, we make use of existing halo model calibrations for the luminosity- \citep{Guo_et_al_2015} and colour-dependent clustering \citep{ss09,pkhp15,pp17b} of the optical SDSS galaxies and prescribe a scaling relation between the optical and \Hi\ properties. We focus on the massive end of the \Hi\ galaxy population which is expected to have substantial overlap with the optically faint end of the galaxy population observed by SDSS. Indeed, cross-matching analyses have found large numbers of galaxies that can be paired between ALFALFA, SDSS and GALEX \citep{Catinella_et_al_2010,toribio+11,maddox+15}.  Our approach to connecting the optical properties with \Hi\ content, however, is statistical, inspired by similar analyses routinely performed in the galaxy cluster community \citep[see, e.g.][]{Mantz_et_al_2010}. In the following, we will assume that the optical properties of ALFALFA galaxies are well-approximated by those measured for SDSS galaxies without evolution, which is reasonable given the small redshift range involved.
\subsection{Setup}
\label{subsec:setup}
\noindent
We assume that \Hi\ mass obeys a scaling relation in terms of optical luminosity $L$ and colour $c$, defined by a mean relation $\avg{\log\mHI|L,c}$ and a scatter $\sigma_{\log\mHI}$ around this mean. For definiteness, we assume that the distribution of \mHI\ at fixed luminosity and colour is Lognormal, so that this mean and width describe the full distribution. We then tie together this scaling with colour-luminosity relations and an optical HOD calibrated by previous analyses to produce a model for the \Hi\ mass function and 2-point correlation function. Schematically, we write a conditional  \Hi-mass function $\phi(\mHI | m)$ in terms of a conditional luminosity-function $\phi(L|m)$, colour-luminosity distribution $p(c|L)$, assumed \Hi-optical scaling distribution $p(\mHI | L,c)$ and the fraction of optical galaxies contributing towards the neutral hydrogen mass of interest $f$ as 
\be
\phi_{g}(\mHI | m) \sim f \times \int \der L\,\phi_{g}(L|m) \int \der c\,p_{g}(c|L)\,p_{g}(\mHI | L,c) \,\, ,
\label{eq:condmHI-schematic}
\ee
which is then used to build a halo model of \Hi\ clustering. In the above expression $g$ denotes the galaxy-type, whether it is a central or a satellite galaxy. We also impose the numerical restriction $m > {\mHI}^{\rm (min)}$ and never consider halos smaller than the smallest \Hi\ content of the galaxies they are expected to host.
 
Appendix~\ref{app:optical} gives the details of our implementation and also discusses the motivation behind some of our technical choices (e.g., we assume that the colour-luminosity relation and \Hi-optical scaling are independent of halo mass $m$).
After exploring various choices for the functional dependence of the mean relation on luminosity and colour, we finally settled on
\begin{align}
\avg{\log\mHI|L,c} &= \overline{\log\mHI} -\alpha_{\rm L}\left(M_r-M_r^{\rm max}\right) \notag\\
&\ph{\mHI}
- \alpha_{\rm c}\log\left[4.66c-1.36c^2-1.108\right]\,,
\label{eq:HIoptscaling}
\end{align}
where we use the SDSS ${}^{0.1}r$-band absolute magnitude $M_r\equiv M_{{}^{0.1}r}-5\log(h)$ in place of luminosity, and set $c={}^{0.1}(g-r)$ to be the colour. $M_r^{\rm max}$ is a luminosity threshold that we discuss later. The sign convention for $\alpha_{\rm L}$ and $\alpha_{\rm c}$ is such that $\alpha_{\rm L}>0$ implies a positive correlation between \Hi\ mass and luminosity and $\alpha_{\rm c}>0$ places more \Hi\ in blue galaxies.
The motivation for the complicated-looking logarithm in \eqn{eq:HIoptscaling} involving colour is the fact that the quantity inside the logarithm gives a good description of the stellar mass-to-light ratio of SDSS galaxies \citep{ww12,pkhp15}, so that
\begin{align}
\log(m_\ast/\Mhsq) &= -0.4\left(M_r-M_{r\odot}\right) \notag\\
&\ph{-0.4}
+ \log\left[4.66c-1.36c^2-1.108\right]\,,
\label{eq:masstolight}
\end{align}
\citep[see equation~7 of][]{pkhp15} where $c\equiv{}^{0.1}(g-r)$ and $M_{r\odot}=4.76$ is the $r$-band absolute magnitude of the Sun. It follows that setting $\alpha_{\rm c}=-2.5\alpha_{\rm L}\equiv-\alpha$ would enforce a power-law scaling between \Hi\ mass and stellar mass, $\mHI\propto m_\ast^\alpha$. The model in \eqn{eq:HIoptscaling} is flexible enough to allow for deviations from a strict dependence of \Hi\ mass on stellar mass, while still stable enough to the dynamic range of optical colours. (Other choices such as linear or quadratic scalings with colour led to unstable results with non-convergent Monte Carlo chains.) 
We then use the MCMC method using the package EMCEE \footnote{\href{http://dfm.io/emcee/current/}{http://dfm.io/emcee/current/}} \citep{Emcee_Mackey_et_al_2013} to constrain the free parameters in our model. In general, all the scaling parameters may be different for centrals and satellites, but the data is not good enough to constrain all of them. So we separate only the parameter $\overline{\log \mHI}$ for centrals and satellites. Other parameters are assumed to be independent of galaxy type.  In this way we are left with a total of six parameters that define the scaling, i.e., $f$, $\overline{\log\mHI}_{\rm cen}$, $\overline{\log\mHI}_{\rm sat}$, $\alpha_{\rm L}$, $\alpha_{\rm c}$ and $\sigma_{\log\mHI}$. The \Hi\ galaxy clustering data that we use is taken from \citet{Guo_et_al_2017}.  For the \Hi\ mass function, we use the binned mass function obtained by taking finite differences of the thresholded mass function as reported in Table~1 of \citet{Guo_et_al_2017}. As compared to the results of \citet{Papastergis_et_al_2013}, this gives us more leverage in the MCMC analysis by probing the shape of the high-mass \Hi\ mass function in some detail. We assume independent errors for these binned measurements.
\subsection{Sample completeness} \label{sec:sample_completeness}
\noindent
Since we are relying on an optical HOD calibrated by \citet{Guo_et_al_2015}, we are restricted to using the luminosity range explored by those authors in their clustering measurements and HOD analysis, which was $M_r<-18.0$. Our luminosity threshold in all integrals is therefore set to $M_r^{\rm max}=-18.0$. Consequently, our analysis must also be restricted to the massive end of the \Hi\ distribution, since we expect at least some positive correlation between luminosity and \Hi\ mass. Exactly which threshold of \mHI\ we can reliably probe, however, cannot be decided \emph{a priori} since the scaling between \Hi\ mass and optical properties that would give us this information is the very quantity we are trying to constrain. To break this circularity and produce stable and converged results, we proceed as follows.
Let us assume that the true scaling is unique, i.e., independent of the \mHI\ threshold. We ran our MCMC chains for different thresholds and found the following:
\begin{itemize}
\item  Chains with the \Hi\ mass threshold $\log [\mHI/\Msun] > x, \forall x<9.8$ do not converge with $M_r^{\rm max} = - 18.0$.
\item Chains with \Hi\ mass threshold $\log [\mHI/\Msun]> x, \forall x \geq 9.8$ converge well for $M_r^{\rm max} = -18.0$.
\item Chains with $\log [\mHI/\Msun]>9.8$ do not converge for higher magnitude thresholds like $M_r^{\rm max} = -19.0$.
\end{itemize}
Depending on these results, we choose $\log [\mHI/\Msun] > 9.8$ as the completeness cut in our \Hi\ mass range. 
%---------------------
\begin{figure*}
\centering
\includegraphics[scale=0.5]{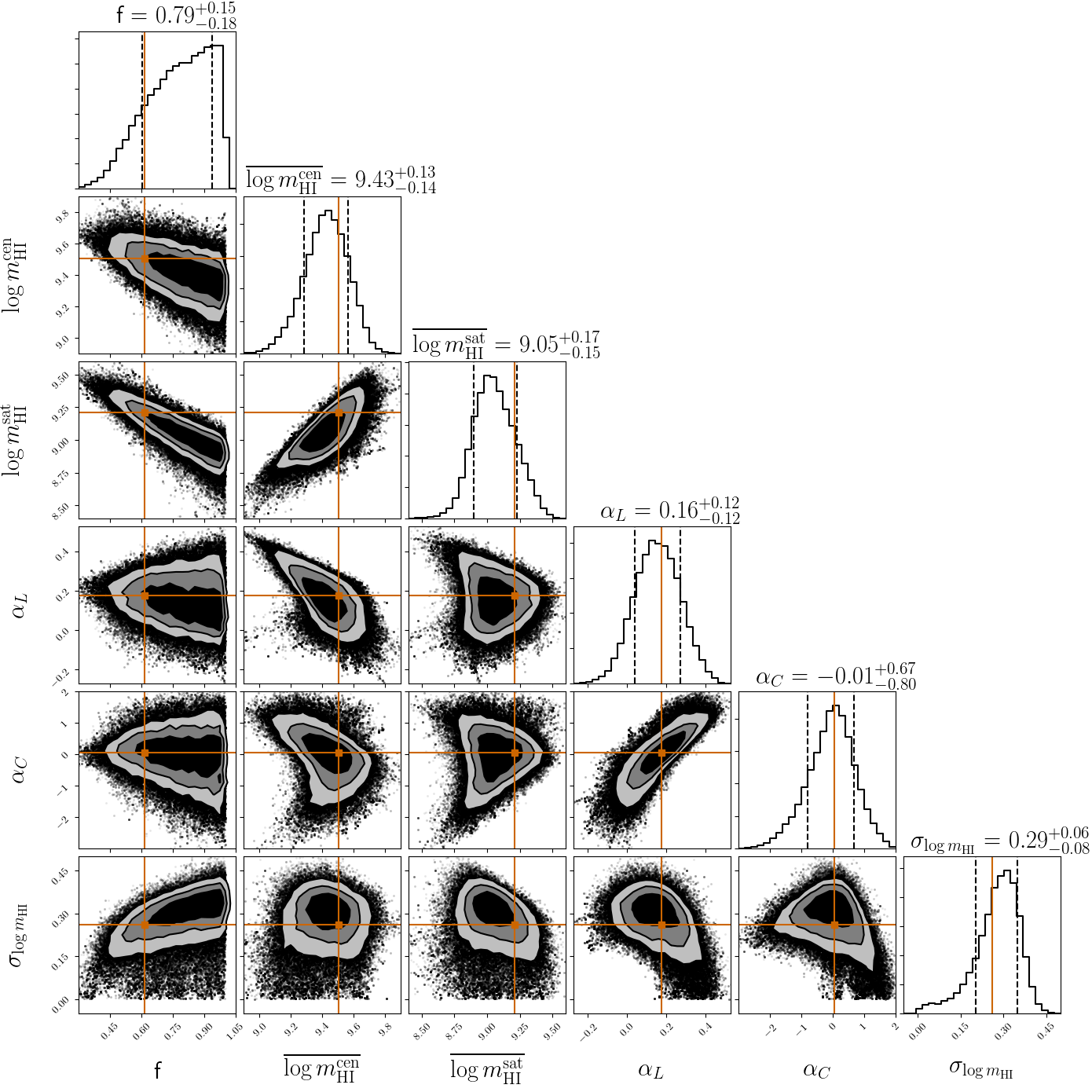} 
\caption{This plot shows the 1, 1.5 and 2-$\sigma$ parameter contours, and the corresponding marginalised 1-D distributions, of the six parameters in our scaling relation analysis (equation~\ref{eq:HIoptscaling}; see also Appendix~\ref{app:optical}) obtained for the combined analysis of the two thresholds $\log [\mHI/\Msun]>9.8$ and $\log [\mHI/\Msun]>10.2$. The orange horizontal and vertical lines denote the best-fit values of the parameters. The values above the 1-D distributions indicate the median and $\pm 1$-$\sigma$ range (i.e., the 16 th and 84 th percentiles) of the respective parameter.}
\label{fig:contour_mHIscaling} 
\end{figure*}
\begin{table*}
\centering
  \begin{tabular}{c c c c c c c c}
 $\log [\mHI/\Msun]$ & $f$  & $\overline{\log\mHI}_{\rm cen}$ & $\overline{\log\mHI}_{\rm sat}$ & $\alpha_{\rm L}$ & $\alpha_{\rm c}$ & $\sigma_{\log\mHI}$ & $\chi^2/ {\rm dof}$ \\
\hline  \hline  
 9.8 & $0.61(0.79^{0.15}_{-0.18})$  & $9.50(9.43^{+0.14}_{-0.14})$ & $9.21(9.05^{+0.17}_{-0.15})$ & $0.18(0.16^{+0.12}_{-0.12})$ & $0.05(0.00^{+0.68}_{-0.80})$ & $0.26(0.29^{+0.06}_{-0.08})$ & 16.43/22 \\
\hline  
  \end{tabular}
  \caption{Summary of results for the scaling relation analysis based on \eqn{eq:HIoptscaling} (see also Appendix~\ref{app:optical}). The columns refer to the \mHI\ threshold used in the analysis, followed by the best-fit (outside parentheses) and median $\pm1\sigma$ values (in parentheses) of the six parameters $f$, $\overline{\log\mHI}_{\rm cen}$, $\overline{\log\mHI}_{\rm sat}$, $\alpha_{\rm L}$, $\alpha_{\rm c}$, $\sigma_{\log\mHI}$, and finally the total Chi-squared and number of degrees of freedom (dof) in the problem. The parameters $\overline{\log\mHI}_{\rm cen}$ and $\overline{\log\mHI}_{\rm sat}$ have units of $\log(\Msun)$. See text for a discussion.}
\label{tab:scaling_relation_parameters}
\end{table*}
\subsection{Results}
\begin{figure}
\centering
\includegraphics[width=0.48\textwidth]{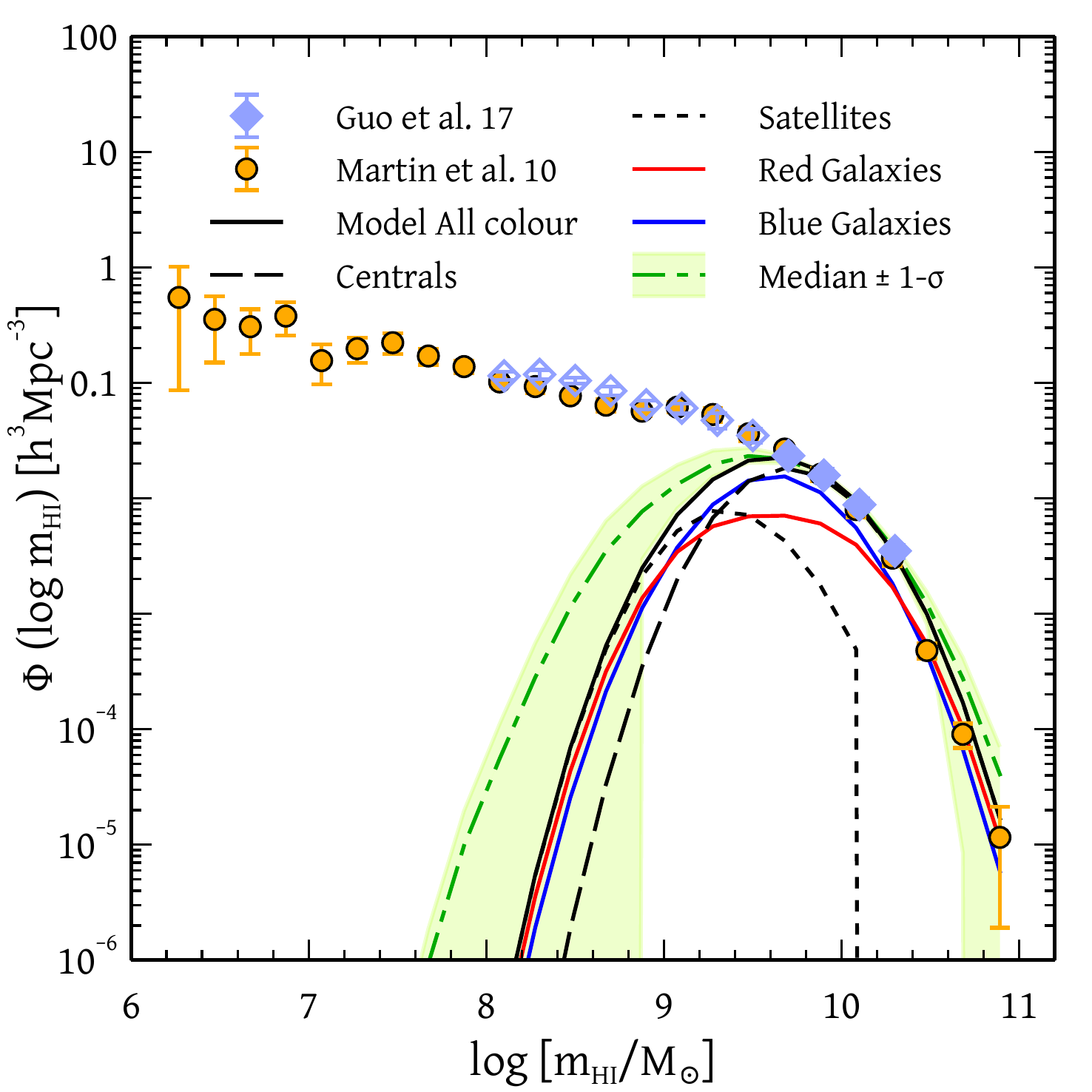} 
\caption{The \Hi\ mass function. The orange circles with error bars are measurements from the $40 \%$ complete ALFALFA survey \citep{Martin_et_al_2010} and the purple diamonds with errorbars are the measurements from $\alpha$-$70$ survey, obtained after taking finite differences of thresholded abundance measurement from Table~1 of \citet{Guo_et_al_2017} (only the four solid purple points were used to constrain our model). The solid black curve shows the mass function computed using our best-fit scaling relation constrained using abundance and clustering data for $\log[\mHI/\Msun]>9.8$ and $\log[\mHI/\Msun]>10.2$ simultaneously. The turn over at the low \Hi\ mass end is due to the incompleteness of the optical HOD based on SDSS data used in our scaling relation. The dashed (dotted) black curve shows the contributions of central (satellite) galaxies using the same best-fit scaling relation. The green dot-dashed curve with the error-band shows the median and $\pm 1$-$\sigma$ error in the \Hi\ mass function as obtained from our distribution of parameters in the MCMC chains. The red (blue) solid curve shows the mass function computed using the scaling relation for red (blue) galaxies. Our model predicts that blue galaxies dominate the \Hi\ content for $9.2 < \log[\mHI/\Msun] < 10.2$, in other ranges the trend is reversed.}
\label{fig:mass_function_mHIscaling} 
\end{figure}

\begin{figure}
\includegraphics[width=0.48\textwidth]{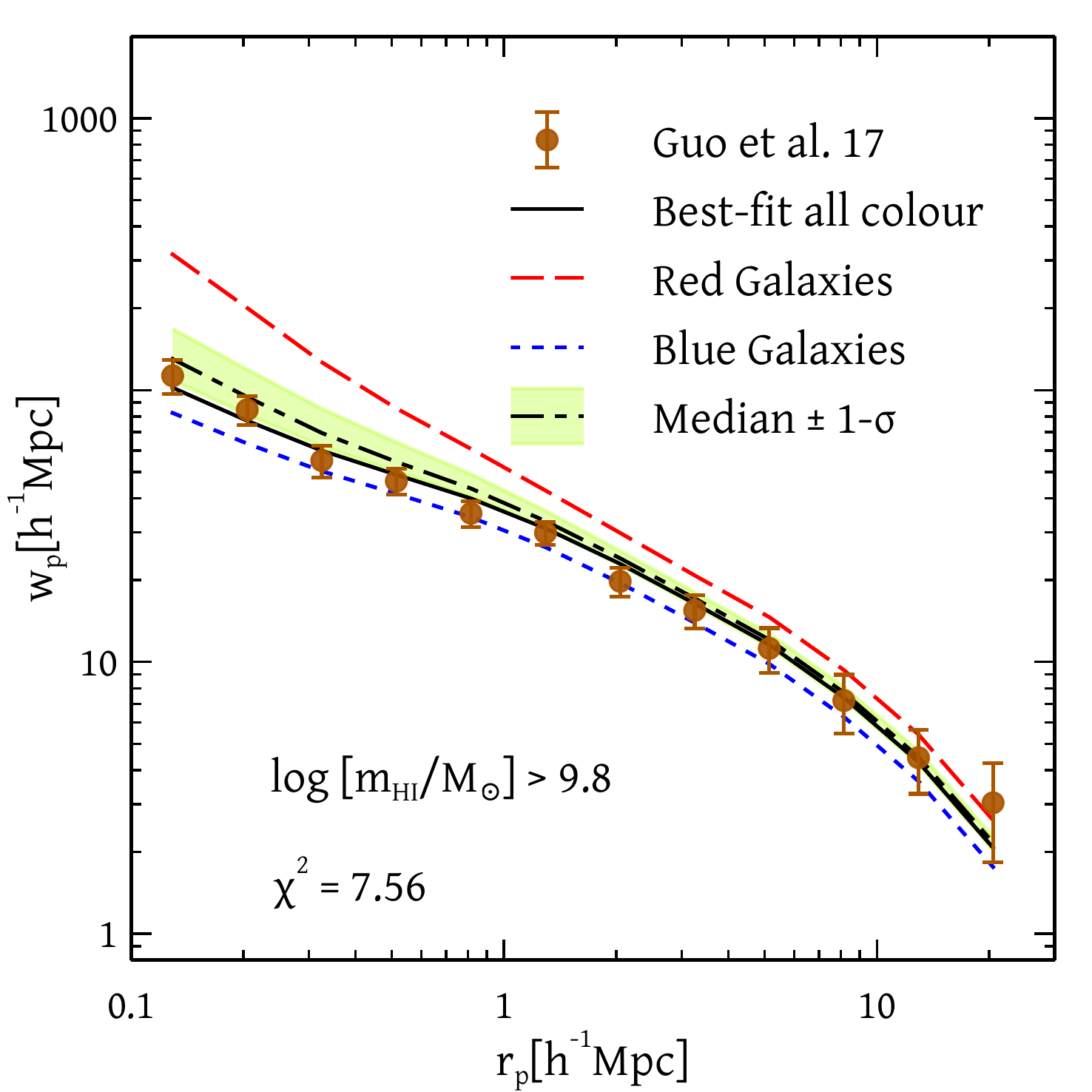}  
\caption{Projected correlation function $w_{\rm p}/r_{\rm p}$ for the threshold $\log [\mHI/\Msun] > 9.8$. The brown points with error bars show the measured correlation function from the ALFALFA survey \citep{Guo_et_al_2017}. The solid black curve shows the correlation function computed using the best-fit scaling relation parameters for all galaxies. The dashed red and dotted blue curves show the same for the optical red and blue galaxies, respectively. The dot-dashed black curve with light green error band shows the median and $\pm 1$-$\sigma$ error in the correlation function obtained from the distribution of parameters in our MCMC chains. The total $\chi^{2}$ obtained from our best-fit scaling relation and using only these data points of correlation measurements has also been quoted.} 
\label{fig:wp_mHIscaling_98} 
\end{figure}

\begin{figure}
\includegraphics[width=0.48\textwidth]{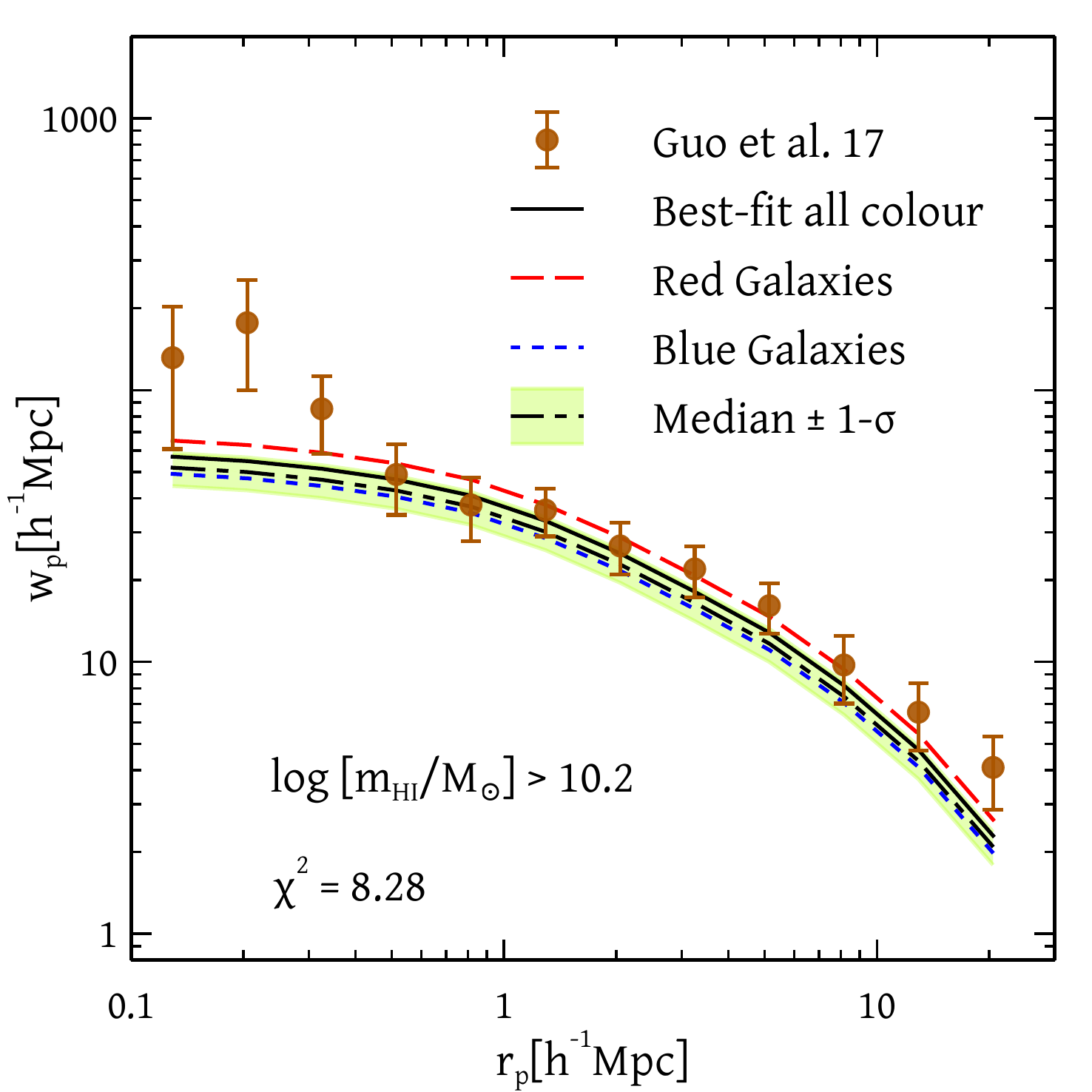}  
\caption{Same as Figure~\ref{fig:wp_mHIscaling_98} but for the threshold $\log [\mHI/\Msun] > 10.2$. } 
\label{fig:wp_mHIscaling_102} 
\end{figure}

\noindent
To take into account for the effect of \Hi\ mass dependent clustering we fit the correlation data for the thresholds $\log [\mHI/\Msun] > 9.8$ and $10.2$ simultaneously assuming that the clustering measurements of these two thresholds are independent. This assumption of zero correlation between clustering measurements of these two thresholds are justified because the \Hi\ mass function drops down substantially from $\log [\mHI/\Msun] = 9.8$ to $10.2$.  Our results from the analysis of the parametric HOD fitting which are similar to those found by \citet{Guo_et_al_2017}, suggests that there will be no satellite for $\log [\mHI/\Msun] > 10.2$. So we put a hard-cut on satellite population above that \Hi\ mass threshold. 
We find a good fit for our six-parameter model constrained by the \Hi\ mass function and correlation function, with a Chi-squared/dof of $16.43/22$ (see Table~\ref{tab:scaling_relation_parameters} for the best-fit parameter values). The corresponding likelihood contours are shown in Figure~\ref{fig:contour_mHIscaling}. We see that the best-fit value of $\alpha_{\rm L}$ corresponds to a small but positive correlation between \Hi\ mass and luminosity. The best-fit value of $\alpha_{\rm c}$ implies that there is little explicit dependence of $p(\mHI|L,c)$ relation on colour. To better understand the characteristics of the best-fit model, we show the corresponding \Hi\ mass function in Figure~\ref{fig:mass_function_mHIscaling}, with the contribution of the red and blue galaxies shown separately.  We see that the mass function is dominated by the blue galaxies in the range we have constrained our model. Outside that range, red galaxies dominate the mass function.  The quantities plotted as the `red' and `blue' HI mass functions, are the ones computed using only the red and blue modes of the double-Gaussian distribution in equation~ \eqref{eq:p_c_given_Mr_cen}. So, even though the integral over each mode approximately gives unity due to $\alpha_C \sim 0$, there will be a difference between the red and blue populations due to the non-trivial dependence of the respective red fractions $p(red | L,g)$ on luminosity as given by equations \eqref{eq:p_red_Mr} to \eqref{eq:predcen}. This difference is seen from the plots of figure \ref{fig:mass_function_mHIscaling}, \ref{fig:wp_mHIscaling_98} and  \ref{fig:wp_mHIscaling_102}.   The turn-over of the predicted mass function at small $m_{\rm HI}$ is due to the hard luminosity cut of $M_r<-18.0$ imposed, as we have mentioned before, by the HOD calibration of \citet{Guo_et_al_2015}. We also see that the number of satellite galaxies suddenly drops to zero at $\log [\mHI/\Msun] = 10.2$. This is due to our assumption that there are no satellites beyond that \Hi\ mass scale. 
\begin{figure}
\includegraphics[width=0.48\textwidth]{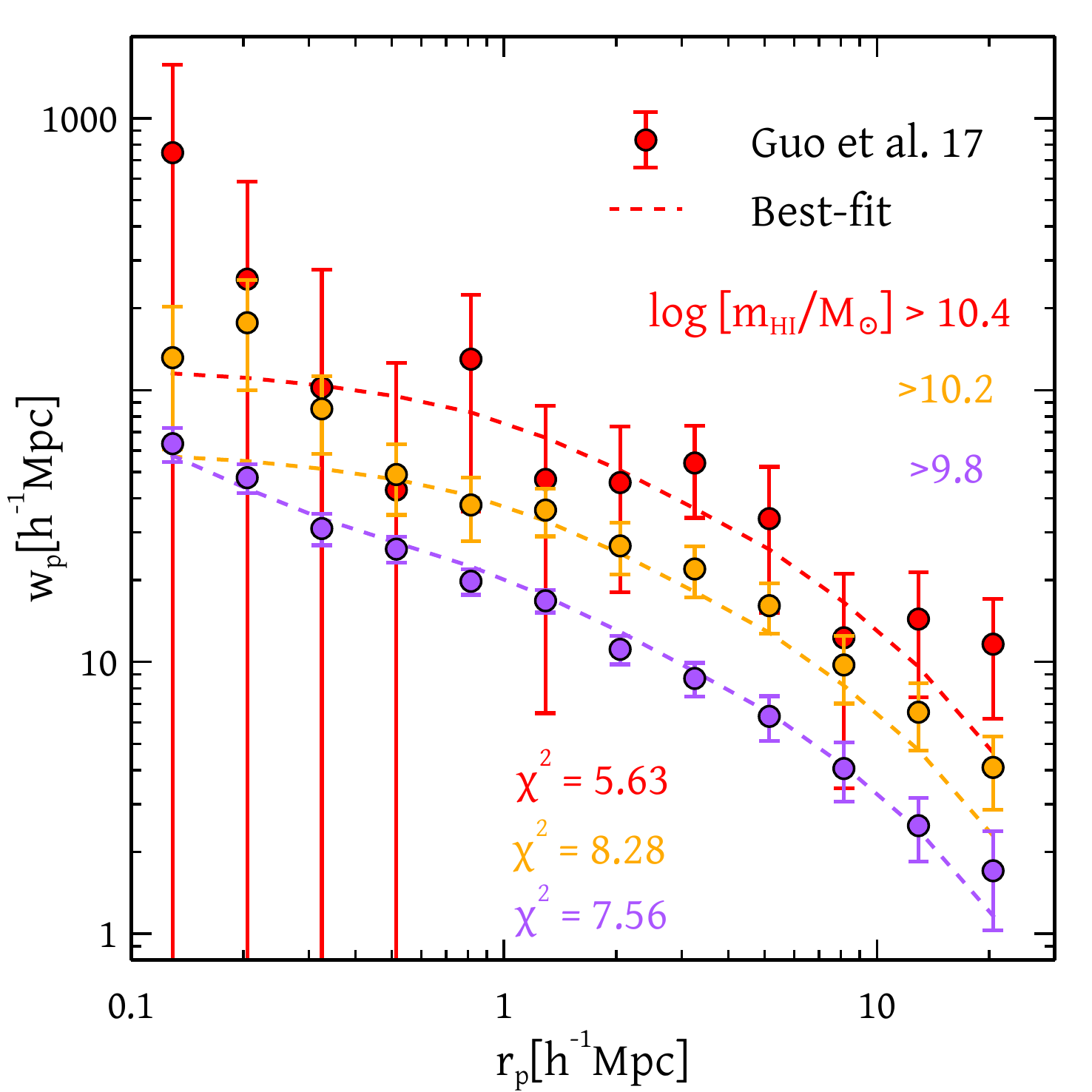}   
\caption{In this figure, we show the projected correlation function of three thresholds $\log [\mHI/\Msun] > 9.8, 10.2$ and $10.4$ altogether. The solid circles with errorbars represent the measurements from \citet{Guo_et_al_2017} using the $\alpha$-$70$ sample and the dotted curves show our model prediction using the best-fit scaling parameters. The correlation functions are separated by $0.25$ dex w.r.t. the middle one and ordering of the larger to smaller thresholds is from top to bottom. We see that our model successfully captures the $\mHI$ dependent clustering strength as observed from the data. A point to note is that the clustering data of only the two thresholds $\log [\mHI/\Msun] > 9.8, 10.2$ were used to constrain our scaling-relation parameters, still this model can predict the clustering for the threshold $\log [\mHI/\Msun] > 10.4$ with quite good accuracy as seen from the plot. The total $\chi^2$ values obtained using the best-fit scaling relation and only the correlation data of different thresholds have also been quoted.} 
\label{fig:wp_mHIscaling_all_together} 
\end{figure}
\begin{figure}
\centering
\includegraphics[width=0.48\textwidth]{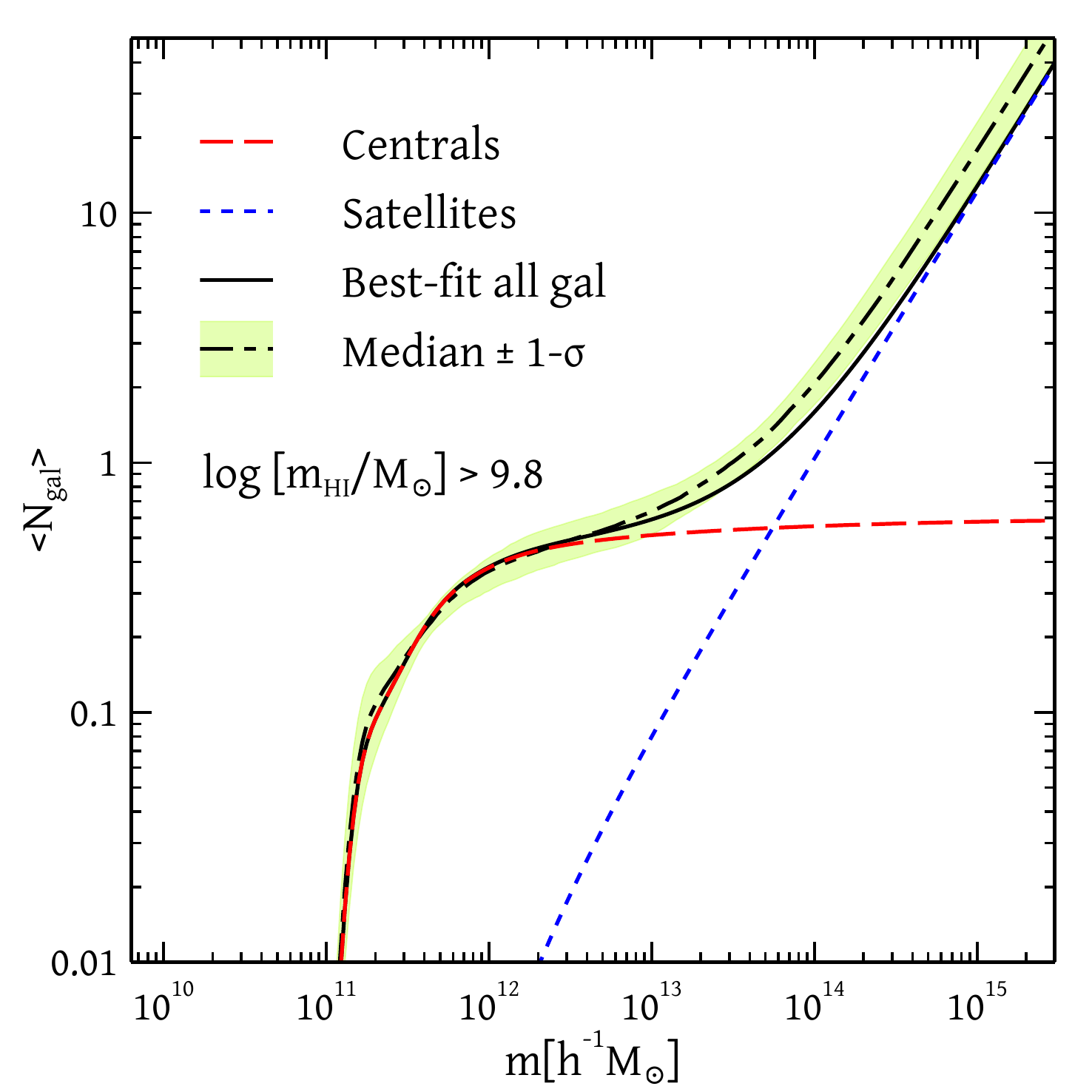} 
\caption{Mean number of galaxies per halo in halos of mass $m$ for the threshold $\log [\mHI/\Msun]>9.8$ as computed from our best-fit scaling relation. The dashed red (dotted blue) line shows the abundances of central (satellite) galaxies as a function of halo mass, and the solid black curve shows the sum of these two. The dotted black line shows the total galaxy abundance using the median values of our parameter chains, and the green band shows the $\pm 1\sigma$ error around the median. We see that there are some satellite galaxies contributing neutral hydrogen for this threshold and their number increases with halo mass.}
\label{fig:population_HOD_98} 
\end{figure}
\begin{figure}
\centering
\includegraphics[width=0.48\textwidth]{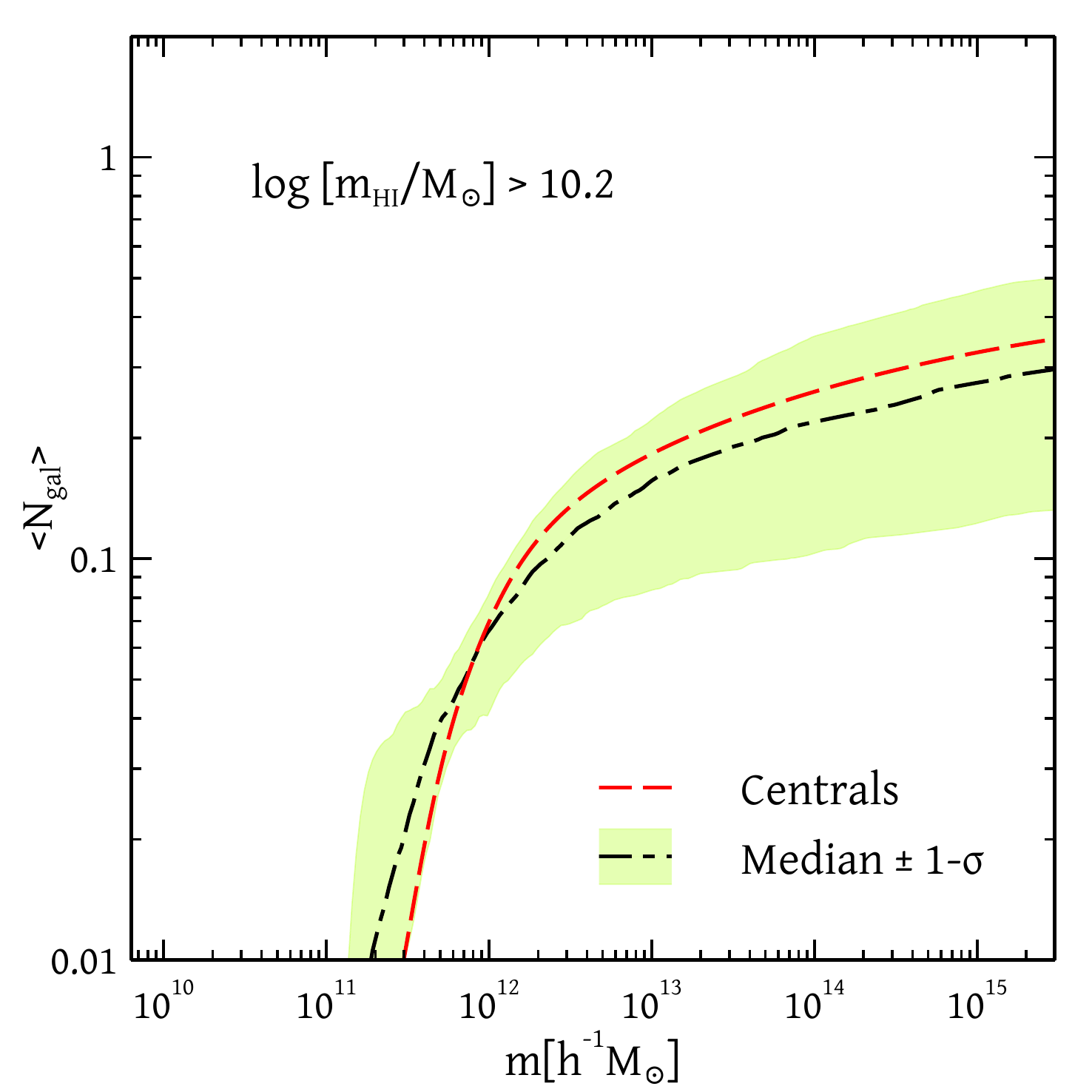} 
\caption{Same as \ref{fig:population_HOD_98}, but for the threshold $\log [\mHI/\Msun]>10.2$. It is to be noted that by construction there is no satellite for this threshold.}
\label{fig:population_HOD_102} 
\end{figure}
\begin{figure}
\includegraphics[width=0.45\textwidth]{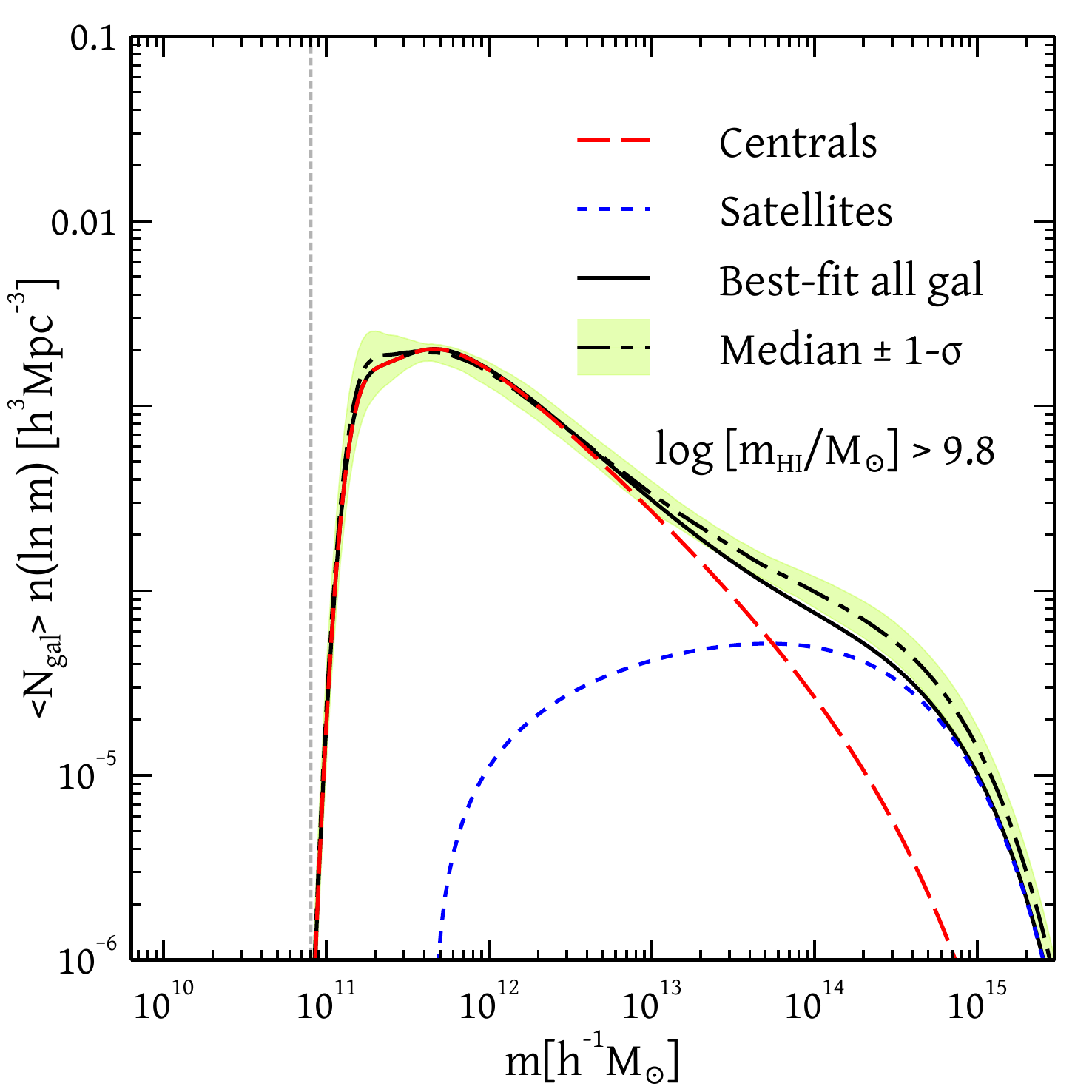}
\caption{Total galaxy number density, defined as the mean number of galaxies per halo multiplied by the halo mass function, as a function of halo mass. The curves and bands are the counterparts of the ones in Figure~\ref{fig:population_HOD_98}. It is clear from this plot that the number density of satellite galaxies contributing towards neutral hydrogen is not very significant. The faint grey dotted line at $m \sim 8\times 10^{10}h^{-1}M_{\odot}$ indicates that the total galaxy number density falls of approximately three orders of magnitude at this halo mass scale.}
\label{fig:HOD_times_hmf_98} 
\end{figure}
The corresponding predictions for the projected \Hi\ correlation function for the best-fit model are shown in Figure~\ref{fig:wp_mHIscaling_98} and \ref{fig:wp_mHIscaling_102} for the two \Hi\ mass thresholds $\log [\mHI/\Msun] > 9.8 $ and $10.2$ respectively. In figure~\ref{fig:wp_mHIscaling_all_together}, we show the projected correlation function for three thresholds separated by $0.25$ dex w.r.t. the middle one. It is to be noted that though the measurements of the threshold $\log [\mHI/\Msun] > 10.4$ were not used to constrain our scaling parameters, our model predicts the clustering of that highest \Hi\ mass threshold with good accuracy. Our model also predicts the clustering amplitude of \Hi\ -hosting red galaxies to be slightly higher than the blue ones \citep[c.f. the colour-dependent clustering of optical galaxies,][]{ss09,Zehavi_et_al_2011}; the data, of course, do not distinguish between these. 
The mean number of galaxies per halo as a function of halo mass -- i.e., an inferred HOD --  obtained from the scaling analysis is shown in Figure~\ref{fig:population_HOD_98} for the threshold $\log [\mHI/\Msun]>9.8$. The contribution from the centrals and the satellites have been shown separately. The median and the uncertainty of all galaxy occupation distribution as obtained from the various parameter sets in our MCMC chains have also been shown. Figure~\ref{fig:population_HOD_102} represents the same information as in  figure~\ref{fig:population_HOD_98} but for the threshold $\log [\mHI/\Msun] > 10.2$. We see that there is no contribution of satellites for this threshold which is by the construction of our model. Figure~\ref{fig:HOD_times_hmf_98} shows an alternative representation of these results, where we have multiplied the HOD of the threshold $\log [\mHI/\Msun] > 9.8$ with the halo mass function \citep[we use the fitting function from][for the latter]{Tinker_2008_mass_function}; the result is the \emph{number density} of galaxies per logarithmic halo mass interval, as a function of halo mass.  We discuss these results further in section~\ref{sec:discuss}.
%
%----------------
%
\section{Discussion}
\label{sec:discuss}
\noindent
In this section, we discuss some implications of our scaling results and compare it with other results from the literature. We also highlight some of the caveats in our analysis and mention a few possibilities for future improvement of our technique.
\subsection{Implications of scaling relation analysis}
\label{subsec:scalingimplications}
\begin{figure}
\centering
\includegraphics[width=0.48\textwidth]{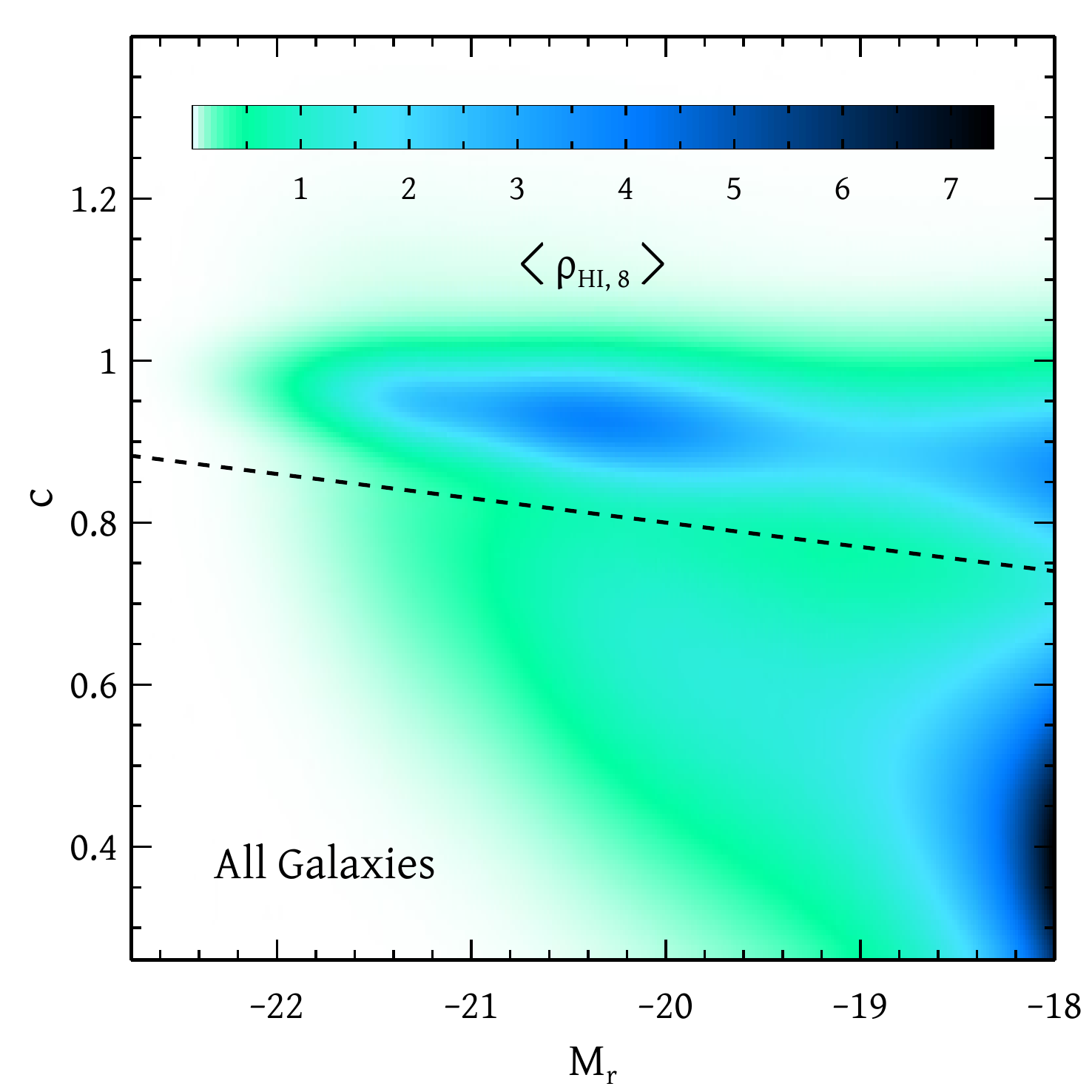} 
\caption{Spatial density of \mHI\ evaluated using our best-fit scaling relation, per unit $c\equiv g-r$ colour, per unit $r$-band magnitude $M_r$, as a function of $c$ and $M_r$. In particular, the colour indicates the value of $\avg{\rho_{\Hi,8}}\equiv\avg{\mHI|c,M_r}p(c|M_r)\phi(M_r)/(10^{8}\Msun h^{3}{\rm Mpc}^{-3})$, where $\phi(M_r)$ and $p(c|M_r)$ are the luminosity function and colour distribution at fixed luminosity, respectively (see Appendix~\ref{app:optical}). We see that a large fraction the neutral hydrogen in SDSS galaxies with $M_r<-18$ in our model resides in faint blue galaxies and some amount in red galaxies too. For comparison, the dashed line shows the relation $c=0.8-0.03(M_r+20)$ which is often used as an empirical separator of red and blue galaxies in optical analyses \citep[see, e.g.,][]{ss09}.} 
\label{fig:2d_histogram_logmHI}
\end{figure}

\noindent
Our scaling relation analysis, which is built on the success of optical HODs calibrated using SDSS DR7 data, allows us to address some interesting questions which the more traditional parametric HOD analysis based only on ALFALFA data cannot. 
For example, our choice of \mHI\ scaling in \eqn{eq:HIoptscaling} allows us to calculate the mean \Hi\ mass density contributed by low-redshift massive galaxies with different luminosities and colours. This is shown in Figure~\ref{fig:2d_histogram_logmHI}, where the coloured region indicates the value of $\avg{\rho_{\Hi,8}}$ defined as the mean \Hi\ mass per unit comoving volume (in units of $10^8\Msun h^3 {\rm Mpc}^{-3}$), per unit optical ($g-r$) colour, per unit $r$-band absolute magnitude, as a function of optical properties. We have used the best-fit values for the scaling relation (Table~\ref{tab:scaling_relation_parameters}) in making this plot, which clearly shows that our model places the bulk of \Hi\ in SDSS galaxies with $M_r<-18$ in faint blue ones, with some amount in red galaxies as well.
\begin{figure}
\centering
\includegraphics[width=0.48\textwidth]{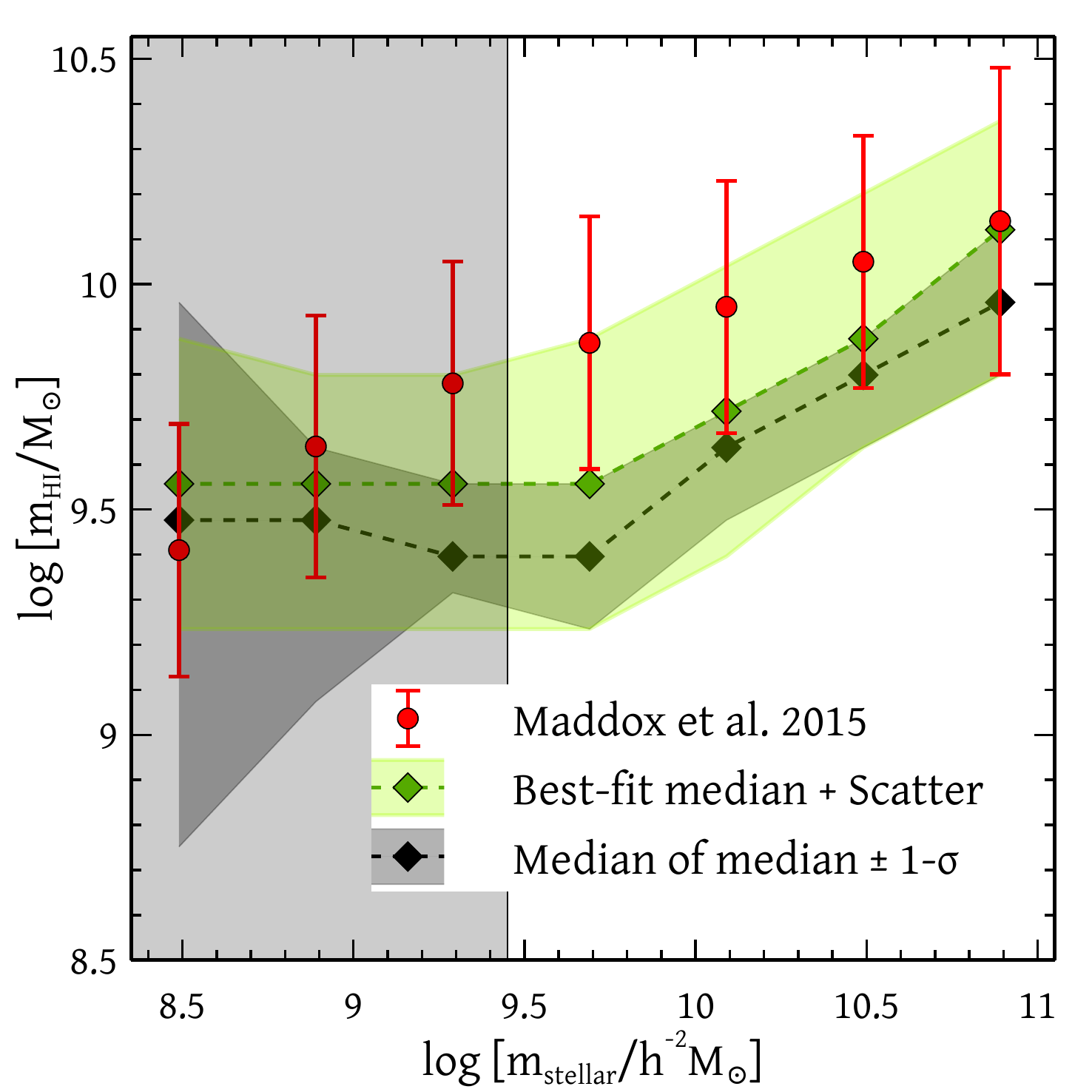} 
\caption{\mHI\ at fixed stellar mass $m_\ast$ predicted by our best-fit scaling relation, as a function of $m_\ast$. The black diamond-shaped points and the dark grey band show the median and $\pm 1$-$\sigma$ width respectively computed from our best-fit scaling relation model. The green diamonds and the light green band show the median of the median and $\pm 1 \sigma$ uncertainty in this relation as computed from various sets of parameters in our MCMC chains. The light grey shaded area on the left indicates values of $m_\ast$ that are affected by incompleteness due to the hard luminosity cut $M_r<-18.0$. For comparison, the red points with error bars show the mean and scatter of the measurements from \citet[][their Table 1]{maddox+15}; note that these were \emph{not} used in constraining our model.} 
\label{fig:mHI_mstar_ratio}
\end{figure}
The analysis above has been observationally motivated, with all variables and relations defined exclusively in terms of directly observable quantities such as \Hi\ mass and optical absolute magnitudes. More physically, one might expect the amount of cold gas represented by \mHI\ to be correlated with the stellar content of galaxies represented by the stellar mass $m_\ast$. As discussed in section~\ref{subsec:setup}, a good description of the correlation between $m_\ast$ and optical observables is given by \eqn{eq:masstolight}. We can, therefore, use our best-fit scaling relation to calculate the mean \Hi\ mass in galaxies containing a fixed amount of stellar mass. This \Hi-to-stellar mass relation has been well-explored in the literature, both observationally \citep[see, e.g.,][]{Catinella_et_al_2010,toribio+11,maddox+15} and theoretically \citep[see, e.g.,][]{dave+13,pbp15}. Figure~\ref{fig:mHI_mstar_ratio} shows our best-fit prediction for this relation, compared with observational constraints from \citet{maddox+15}. The gray shaded area indicates values of $m_\ast$ that are affected by incompleteness due to the hard luminosity cut $M_r<-18.0$, calculated by comparing the stellar mass function predicted by the optical HOD (Appendix~\ref{app:optical}) and \eqn{eq:masstolight} with Schechter-function fits to the measured SDSS stellar mass function from \citet{peng+12}.  The error bars on the \citet{maddox+15} data show the measured scatter of the \mHI\ distribution at fixed $m_\ast$.   Correspondingly, the light green band on the green symbols shows the width of this relation predicted from our model using best-fit scaling relation parameters. We have also shown the median and $\pm 1$-$\sigma$ uncertainty of this prediction arising from different sets of scaling parameters of our MCMC chains as the black diamonds and the dark grey band. We see that our model predicts the slope and scatter of the relation reasonably well, although it underestimates the amplitude of the scaling by about $\sim0.2$-$0.3$dex. We are currently investigating whether this could be connected to selection effects introduced by the matching procedure in the observational samples.
We emphasise that although our scaling relation analysis is restricted to the high \Hi\ mass end, it provides a new statistical method to connect the \Hi\ distribution to optical properties of galaxies, which is complementary to direct observational analyses of matched galaxy samples. A novelty of our method is that, since it is a halo model by construction, it also provides a simultaneous theoretical connection of the \Hi\ and optical properties with the underlying dark matter distribution.  Unlike the semi-analytical models of galaxy formation \citep{Lacey_et_al_2016, Zoldan_et_al_2017, Gonzalez_et_al_2018}, our scaling model does not require any prior assumption about different physical processes involved, the information about all those processes can be extracted later from the statistically constrained values of the parameters in the model. 
   Additionally, the analysis in Appendix~\ref{app:HItohalomass} shows that our best fit model produces physically
meaningful results for the ratio of \Hi\ mass to halo mass
in the majority of halos in which they place \Hi\ .  In section~\ref{sec:conclude}, we mention a few interesting avenues for future work along similar lines.
%-------
\subsection{Caveats}
\noindent
Our scaling relation model could benefit from a complete joint analysis of the optical HOD together with the optical-\Hi\ scaling relation. This is primarily because the colour-luminosity relations we have used in this work \citep[][see Appendix~\ref{app:optical} for details]{pkhp15,pp17b} have all been calibrated separately and independently of the luminosity-dependent HOD of \citet{Guo_et_al_2015}. In particular, the calibration of the satellite red fraction $p({\rm red}|{\rm sat},M_r)$ (equation~\ref{eq:predsat}) taken from \citet{pp17b} is the most susceptible to systematic effects, since it was not based on a rigorous Monte Carlo fit to clustering data \citep[see also][]{ss09}. This aspect of our analysis can, therefore, be improved by simultaneously using clustering information from SDSS and ALFALFA in a joint analysis.
In the same spirit, several ingredients of the analytical model used in this work --  particularly those in the $2$-halo regime, such as our treatment of scale dependence of halo bias, or the fact that we ignore halo exclusion -- are affected by systematic uncertainty \citep[see, e.g.,][for a comprehensive study]{vdb+13} which would result in biases on the parameter values being constrained. 
%This affects both the scaling relation as well as the parametric HOD analysis. 
By comparing with mock galaxy catalogues, \citet{pp17b} had estimated these effects to be of the order of $\sim20\%$ in the predictions of SDSS colour-dependent clustering. These uncertainties can also be largely mitigated by using numerical calibrations of halo clustering based directly on $N$-body simulations \citep{zg16}. We are currently incorporating such calibrations in our MCMC routines, and the results will be presented in forthcoming work (Pahwa et al., in preparation).
\subsection{Predictions for Future Missions} 
\label{sec:prediction}
The upcoming optical surveys like Euclid\footnote{\href{http://sci.esa.int/euclid/}{http://sci.esa.int/euclid/}} and LSST\footnote{\href{https://www.lsst.org/}{https://www.lsst.org/}} are expected to have the sensitivity to observe faint galaxies at higher redshifts. On the other hand, the upcoming radio telescope SKA\footnote{\href{https://www.skatelescope.org/}{https://www.skatelescope.org/}} will be able to observe the faint \Hi\ -21 cm signal arising from farther distances. If we assume that our optical HOD, colour-luminosity relation and the probability of assigning \Hi\ mass based on luminosity and $g-r$ colour remain invariant, then we can estimate the number of galaxies that the optical and \Hi\ surveys will be able to detect in a redshift range and also their clustering properties. If the redshift range covered by a survey is $z_{\rm min} < z < z_{\rm max}$ and if it covers a comoving volume $V$, then we can write
\begin{align}
N_{\rm gal} &= \int \der m \int \der z <N_{\rm gal}(m)> n(m,z) \der V/\der z \,\, , \nonumber \\
b  &= \frac{\int \der m \int \der z <N_{\rm gal}(m)> b(m,z) n(m,z) \der V/\der z}{\int \der m \int \der z <N_{\rm gal}(m)> n(m,z) \der V/\der z} \,\, . \nonumber
\end{align}
In the above expressions, $N_{\rm gal}$ is the number of galaxies within the survey volume, $b$ is their large-scale clustering bias and $<N_{\rm gal}(m)>$ is the average number of galaxies in $m$-halos (halos of mass in the range $m$ to $m + \der m$).  All of these three quantities will be functions of either the luminosity threshold or the \Hi\ mass threshold. For the purpose of calculating predictions for the upcoming surveys, we have assumed that only the halo mass function and halo bias change with redshift, while the HOD $\langle N_{\rm gal}(|m) \rangle$ remains identical to what we have obtained at low redshifts. In the scaling analysis performed in this article, we have used the halo mass function of \citet{Tinker_2008_mass_function} and the scale-dependent halo bias prescription of \citet{Tinker_et_al_2005} while the large-scale halo bias is taken from \citep{Tinker_2010_halo_bias}. However, there exists more recent calibrations of these quantities for the Planck cosmology \citep{Despali_et_al_2016, Comparat_et_al_2017}. In the following, to make the predictions for future missions, we will use the mass function and halo bias calibrations reported by \citet[][see their Table~4]{Comparat_et_al_2017}.
    
Our scaling model that we have constrained in the previous section is complete for $\log [\mHI/\Msun]>9.8$ and $M_r < -18.0$. Euclid visible photometric survey has a $r$-band limiting magnitude of $24.5$ \citep{Synergy_Euclid_SKA_Kitching_et_al_2015, Euclid_study_report_2011}. Similarly LSST has a $r$-band magnitude limit of $24.0$ \citep{LSST_science_book_2009}. So the Euclid (LSST) photometric survey will be able to detect $M_r<-18.0$ galaxies upto redshift $z \sim 0.5 (0.44)$. Both of these photometric surveys will cover an area of $20000 {\rm deg}^2$ in the sky. The surveys with the SKA are still being planned. It has been claimed that the SKA shallow, medium deep and deep surveys  will be able to detect $\log [\mHI/\Msun] \sim 9.8$ up to redshifts $0.27$, $0.37$ and $0.69$  with a sky-areas of $10000$, $200$ and $150$ $\rm deg^2$ respectively \citep{SKA_spec_Blyth_et_al_2015}. So in the region of overlap in redshift and sky-area, Euclid (LSST) will detect a total of $\sim 1.6 \times 10^7$ optical galaxies with $M_r<-18.0$ with a corresponding clustering bias of $b_{\rm opt} \sim 1.0$. Similarly SKA will detect a total of $\sim 2.7 \times 10^6$ \Hi\ galaxies with $\log [\mHI/\Msun] > 9.8$ with a slightly smaller clustering bias of $b_{\Hi} \sim 0.8$. The values above for the expected counts are more than two orders of magnitude larger than those have been used in this work. The corresponding values for the clustering bias will be relevant for forecasts of the detectability, using multiple tracers, of ultra-large-scale relativistic effects and primordial non-Gaussianity \citep[see, e.g.][]{Alonso_Ferreira_2015}.  
\section{Conclusion}
\label{sec:conclude}
\noindent
The study of \Hi\ in galaxies is important because of its connection with star formation and galaxy evolution. In particular, connecting these galaxies to the properties of the host dark matter haloes can be of great help in comparing theoretical models with observations. In this work, we have used a novel approach, based on an assumed scaling relation between \Hi\ content and optical properties of galaxies -- to model the mass function and clustering properties of \Hi\ galaxies as measured in the ALFALFA $\alpha.70$ survey.
Our main findings are:
\begin{itemize}
\item 
At intermediate \Hi\ masses ($ 10^{9.8} \lesssim \mHI\lesssim10^{10.2}\Msun$), the clustering data demand that some amount of \Hi\ be placed in satellite galaxies (Figure~\ref{fig:population_HOD_98}), with the number of satellites increasing as a function of halo mass.
\item At very high masses $\mHI > 10^{10.2}\Msun$, most \Hi\ must be in central galaxies (Figure~\ref{fig:population_HOD_102}). 
\item The analysis in section~\ref{sec:scaling} for the sample $\log [\mHI/\Msun] > 9.8$ implies that this \Hi\ is primarily in \emph{optically faint blue galaxies} (Figure~\ref{fig:2d_histogram_logmHI}), which is consistent with observations from matched samples and expectations from galaxy formation models.
\end{itemize}
Although our model has been restricted to the simplest `halo mass only' flavour of HOD analysis, it can easily be extended to include effects such as `galactic conformity' \citep{weinmann+06}, in which satellite galaxies `know' about the star formation properties (or optical colour) of their central, with blue centrals being preferentially surrounded by blue satellites at fixed halo mass. Our optical-\Hi\ scaling analysis can ride on previously developed formalism \citep{pkhp15,pp17b} which allows for galactic conformity to be included in the analysis in a tunable manner. The addition of \Hi\ abundance and clustering data will add new leverage in the problem and is expected to be very useful in constraining the nature of galactic conformity, particularly at large separations \citep{kauffmann+13,hearin+15a}. We will return to these questions in future work. 
\section*{Acknowledgments}
We are grateful to E. Papastergis for kindly providing us with the covariance matrix estimates for the clustering data used in an earlier version of this work. We thank Neeraj Gupta for useful discussions in the initial stages of this work, and R. Srianand for discussions throughout. We also thank H. Guo for making his 2PCF and covariance matrix measurements of $\alpha.70$ sample public and for his valuable comments on an earlier draft. We sincerely thank the anonymous referee for useful suggestions towards the improvement of our analysis. We gratefully acknowledge the use of high performance computing facilities at IUCAA, Pune (http://hpc.iucaa.in). This work used the open source computing packages \textsc{NumPy}  \citep[][http://www.numpy.org]{vanderwalt-numpy}, \textsc{SciPy} \citep{scipy_package} and the plotting softwares \textsc{Veusz} (https://veusz.github.io/) and corner.py \citep{corner}. NP acknowledges the financial support from the Council of Scientific and Industrial Research (CSIR), India as a Shyama Prasad Mukherjee  Fellow. The research of AP is supported by the Associateship Scheme of ICTP, Trieste and the Ramanujan Fellowship awarded by the Department of Science and Technology, Government of India.
\bibliography{HI_HOD_bibliography}
\appendix
\section{Implementation details of scaling analysis}
\label{app:optical}
In this Appendix, we give the details of our implementation of the analytical model describing the abundance and clustering of \Hi\ galaxies via a suitable scaling between \Hi\ mass and optical properties. 
In the following we will consistently denote SDSS galaxy colour as $c\equiv{}^{0.1}(g-r)$ and galaxy type (whether central or satellite) will be denoted by $g$.  Since in principle the scaling relations of centrals and satellites can be different, \eqn{eq:condmHI-schematic} can be explicitly rewritten as,
\begin{align}
\phi_{\rm cen}(\log \mHI |m) = f & \times \int \der M_r \, \phi_{\rm cen}(M_r|m) \int \der c \, p(c|M_r, \text{cen}) \notag \\
& \times p(\log \mHI |M_r, c, \text{cen}) \,\,, \label{eq:fcen_scaling} \\
\phi_{\rm sat}(\log \mHI |m) = f &\times \int \der M_r \, \phi_{\rm sat}(M_r|m) \int \der c \,  p(c|M_r, \text{sat}) \notag \\
& \times p(\log \mHI |M_r, c, \text{sat}) \,\,. \label{eq:Script_N_scaling}
\end{align}
The quantities $\phi_{\rm cen}(M_r|m)$ and $\phi_{\rm sat}(M_r|m)$ are determined by the optical HOD analysis of \citet{Guo_et_al_2015} for different magnitude thresholds. We fit the parameters with suitable functional forms and take finite differences to get the quantities in different magnitude bins. The reader may note that we have not accounted for any halo mass dependence in the remaining functions $p(c|M_r, g)$ and $p(\log \mHI |M_r, g)$, which respectively refer to the colour distribution of galaxies at fixed luminosity and the distribution of \Hi\ in galaxies of fixed luminosity and colour. We discuss the reasons for this below.
The observed colour-luminosity distribution $p(c|M_r)$ of SDSS galaxies has a well-known bimodality (ultimately related to the star-formation properties of galaxies) which can be modelled as a double-Gaussian in colour at fixed luminosity \citep[see, e.g.,][]{Baldry_et_al_2004}, which we can write as
\begin{align}
p(c|M_r) &=  p({\text{red}} | M_r) p_{\text{red}}(c|M_r) \notag\\
&\ph{p(red)}
+ \left( 1 - p({\text{red}}|M_r) \right)\,p_{\text{blue}}(c|M_r) \,, \label{eq:p_c_given_Mr}
\end{align}
where $p({\text{red}} | M_r)$ is the `red fraction' of galaxies with luminosity $M_r$, and $p_{\rm red/blue}(c|M_r)$ are the red/blue modes of the distribution.
When constructing a halo model that includes colour information, one must additionally model the colour distribution of satellites and centrals. \citet{ss09} showed that the data allow us to think of each of these as simply weighted versions of the same two Gaussian modes used for modelling the observed colour distribution. Since the all-galaxy sample leads to very accurate double-Gaussian fits, this means that the uncertainty in the individual shapes of the central and satellite colour distributions is relegated to two functions $p({\rm red}|m,M_r,{\rm sat/cen})$ that describe the red fraction of satellites/centrals of luminosity $M_r$ in $m$-halos. These functions can be constrained by comparing with the colour-dependence of clustering in SDSS. 
As \citet{ss09} further demonstrated using SDSS DR5 data \citep[this has also been confirmed by][for SDSS DR7 data]{pkhp15}, a final simplification is that these satellite/central red fractions can be safely assumed to be independent of halo mass. This also implies that the full colour distributions of centrals (and satellites) are independent of halo mass. Since the centrals and satellites must add up to the full galaxy population whose (halo mass independent) colour distribution is directly observed, this reduces the problem to calibrating a single function of luminosity, which can be chosen to be the satellite red fraction $p({\rm red}|M_r, \text{sat})$.
In detail then, the quantity $p(c|M_r, g)$ we require in our model is built as follows \citep[see also Appendix A of][]{pkhp15}:
\begin{align}
p(c|M_r, g) =  p&({\text{red}} | M_r, g) p_{\text{red}}(c|M_r) \nonumber \\ 
& + \left[ 1 - p({\text{red}}|M_r, g) \right]p_{\text{blue}}(c|M_r) \,\, . 
\label{eq:p_c_given_Mr_cen}
\end{align}
The distributions $p_{\text{red}}(c|M_r)$ and $p_{\text{blue}}(c|M_r)$ are Gaussians with mean $\langle c|M_r \rangle_{\text{red}}$ and $ \langle c|M_r \rangle_{\text{blue}}$ and standard deviation $\sigma_{\text{red}}(M_r)$ and $\sigma_{\text{blue}}(M_r)$. These were calibrated by \citet{pkhp15} to have the forms
\begin{align}
\langle c | M_r \rangle_{\rm red} &= 0.9050 - 0.0257 (M_r + 19.5) \nonumber \\
\langle c | M_r \rangle_{\rm blue} &= 0.575 - 0.126 (M_r + 19.5) \nonumber \\
\sigma_{\rm red} (M_r ) &= 0.0519 + 0.0085 (M_r + 19.5) \nonumber \\
\sigma_{\rm blue} (M_r ) &= 0.150 + 0.015 (M_r + 19.5) \label{eq:Mean_spread_color_mag} \,\, .
\end{align}
We will also need expressions for the all-galaxy red fraction $p(\text{red}|M_r)$ and the satellite red fraction $p(\text{red}|M_r,{\rm sat})$. The former was calibrated by \citet{pkhp15} to have the form
\be
p({\rm red} | M_r) = 0.423 - 0.175 (M_r + 19.5) \,, 
\label{eq:p_red_Mr}
\ee
while \citet{pp17b} showed that the satellite red fraction given by
\be
p({\rm red}|M_r, {\rm sat})  = 1.0 - 0.33 \left[1 + \tanh \left( \frac{M_r + 19.25}{2.1} \right) \right] \,\, ,
\label{eq:predsat}
\ee 
gives a good description of SDSS colour-dependent clustering, although this conclusion was not based on a rigorous statistical analysis. We discuss possible improvements on calibrating this ingredient in the main text.
Putting things together, the central red fraction $p({\text{red}}|M_r, \text{cen})$ implicitly appearing in \eqn{eq:p_c_given_Mr_cen} can be written as
\begin{align}
%% p({\rm red} | M_r, {\rm cen}) & = \left[ p({\rm red}| M_r) - p({\rm red}|M_r, {\rm sat}) \right. \nonumber \\
%% & \left. \times \bar{p}({\rm sat}|M_r) \right]/\bar{p}({\rm cen}|M_r) \,\, , 
p({\rm red}|M_r,{\rm cen}) &= \frac1{\bar{p}({\rm cen}|M_r)}\bigg[p({\rm red}| M_r)\notag\\ 
&\ph{p(red)}
- \bar{p}({\rm sat}|M_r)\,p({\rm red}|M_r, {\rm sat})\bigg]
\label{eq:predcen}
\end{align}
where 
\be
 \bar{p}({\rm cen}|M_r) = \frac{\int \der m\,n(m) \phi_{\rm cen} (M_r|m) }{\int \der m\,n(m)[ \phi_{\rm cen} (M_r|m) + \phi_{\rm sat} (M_r|m)]} 
\ee
and $\bar{p}({\rm sat | M_r}) = 1 - \bar{p}({\rm cen}|M_r)$. 
As regards the halo mass dependence of the \Hi-optical scaling distribution $p(\log\mHI|m,M_r,c, g)$, the quality of the clustering data forces us to minimise the number of free parameters in the model. For simplicity, therefore, we assume this distribution to be independent of halo mass. Inspired by the near-symmetric distributions for $\log\mHI$ found by matched analyses such as \citet{maddox+15} or \citet{Catinella_et_al_2010}, we model $p(\log \mHI|M_r,c, g)$ to be Gaussian with mean $\avg{\log \mHI | Mr, c, g}$ (c.f. equation~\ref{eq:HIoptscaling}) and standard deviation $\sigma_{\log \mHI}$ as discussed in section~\ref{sec:scaling}. The good quality of the fit we achieve (see Table~\ref{tab:scaling_relation_parameters}) is a \emph{post hoc} justification that this model is indeed acceptable.
Finally, \eqn{eq:fcen_scaling} and \eqref{eq:Script_N_scaling} describes the scaling relation for galaxies in bins of \Hi\ mass. To get the HOD $\phi_{\rm cen}(>\mHI|m)$, $\phi_{\rm sat}(>\mHI|m)$ for thresholds of \Hi\ mass (this is essential since the clustering data is only available for such thresholds), we need the quantity $p(> \log \mHI|M_r, c, g)$ which is simply given by the expression, 
\begin{align}
&p(>\log \mHI |M_r, c, g) \notag\\
&=  \frac{1}{2} \left[ \erf{\frac{\log \mHI^{\rm cut, g} - \langle \log \mHI | M_r , c, g \rangle}{\sqrt{2}\sigma_{\log \mHI}}} \right. \notag \\
&\left. - \erf{\frac{\log \mHI - \langle \log \mHI | M_r , c, g \rangle}{\sqrt{2}\sigma_{\log \mHI}}} \right] \,,
\end{align}
where $\erf{x} = (2/\sqrt{\pi})\int_0^x\der y\,{\rm e}^{-y^2}$ is the error function. Once we have the two quantities $\phi_{\rm cen}(>\mHI|m)$ and $\phi_{\rm sat}(>\mHI|m)$, the two-point correlation function can be easily computed via a few simple steps \citep{Cooray_Sheth_2002}.
Given a scaling relation (i.e., values for the six parameters in equation~\ref{eq:HIoptscaling}), we can also use \eqn{eq:fcen_scaling} and \eqref{eq:Script_N_scaling} to calculate the binned \Hi\ mass function by averaging over halo mass,
\begin{align}
\phi(\log \mHI) = \int \der m\,n(m)\, [\phi_{\rm cen}(\log \mHI |m) + \phi_{\rm sat}(\log \mHI |m)] \,.
\label{eq:HI_mass_function}
\end{align}
\section{HI-to-halo mass relation}
\label{app:HItohalomass}
\noindent
\citet{Guo_et_al_2017} have reported that a HOD analysis of their data-set led to unphysical results, with \Hi\ being placed in halos that were too small, giving $\mHI/m$ ratios larger than unity in some cases. In this context, since scaling analysis have produced good fits using measurements from \citet{Guo_et_al_2017}, it is worth asking whether these fits are physically meaningful. To this end, in this Appendix, we explore the distribution of \mHI\ in halos of fixed mass $m$ as implied by our best-fit scaling relation. In particular, we are interested in asking whether a substantial number of halos end up containing `too much' \Hi\ for their halo mass, which would result in the problem reported by \citet{Guo_et_al_2017}.
\begin{figure}
\centering
\includegraphics[width=0.48\textwidth]{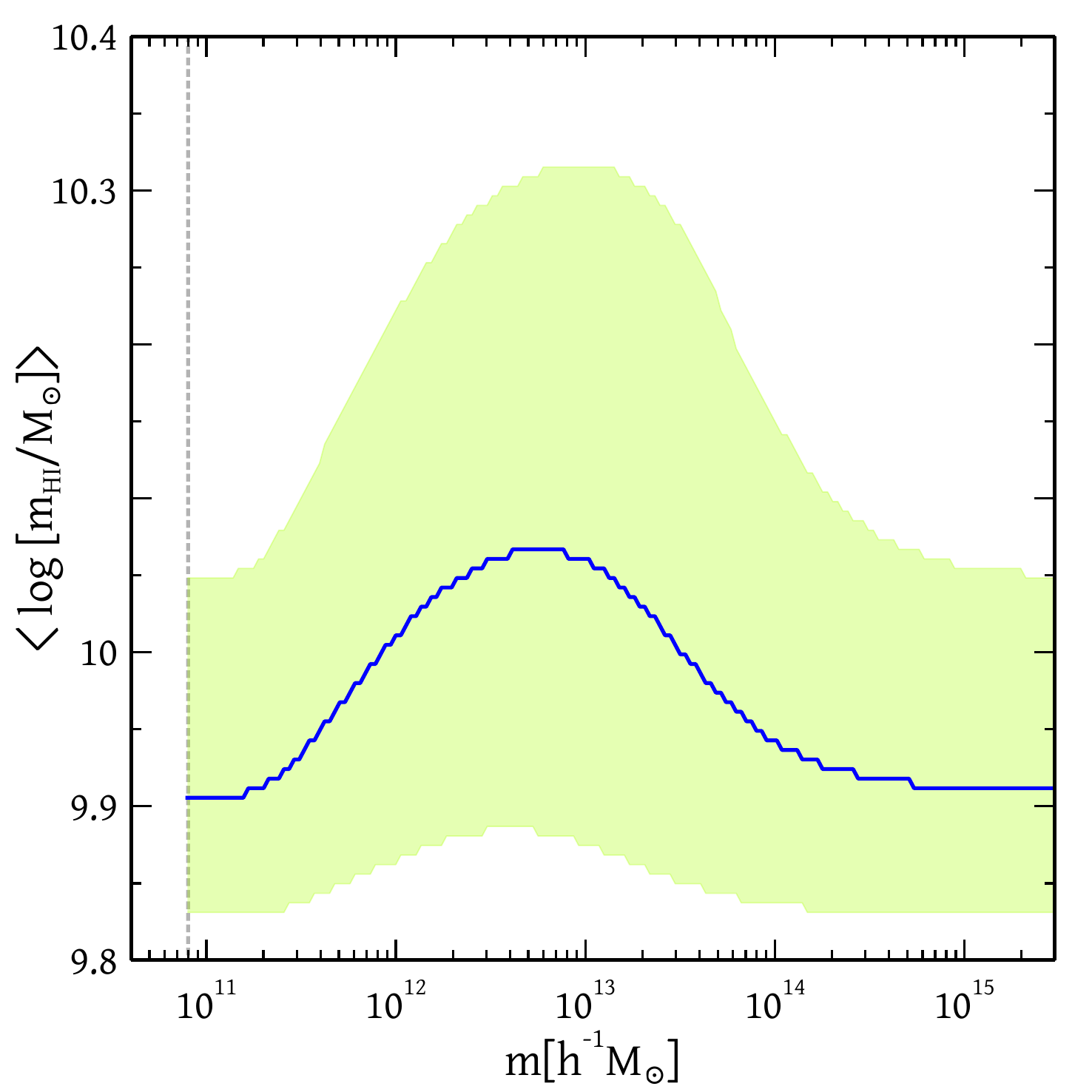} 
\caption{Median (solid curve) and $\pm1\sigma$ range (light green shaded band) of the distribution $p(\log\mHI|m)$ (equation~\ref{eq:plogmHI}). Results are shown for the best-fit scaling relation of section~\ref{sec:scaling} and therefore refer to SDSS galaxies with $M_r<-18.0$. See text for a discussion. The faint grey dotted line on the left corresponds to the same halo mass scale as mentioned in the caption of the figure \ref{fig:HOD_times_hmf_98}. We ignore the distribution below that halo mass scale since the number density of those halos are very small.}
\label{fig:plogmHI_scaling} 
\end{figure}
Ideally, we would therefore like to explore the full probability distribution $p(\log\mHI|m)$ in $m$-halos occupied by \Hi, as a function of halo mass. This is in fact possible in our scaling relation analysis where we can calculate this function as
\be
p(\log\mHI|m) = \frac{\phi_{\rm tot}(\log \mHI |m)}{\int\der\log\mHI\,\phi_{\rm tot}(\log \mHI |m)} \,\, ,
\label{eq:plogmHI}
\ee
where $\phi_{\rm tot}(\log \mHI |m) = \phi_{\rm cen }(\log \mHI |m) + \phi_{\rm sat}(\log \mHI |m)$. Figure~\ref{fig:plogmHI_scaling} shows the median and $\pm1\sigma$ range (i.e., the $16^{\rm th}$ and $84^{\rm th}$ percentiles) of this distribution. We see that  the distribution is very well behaved, with $\mHI/m$ predicted to be $<0.15$ everywhere above the halo mass scale $m > 8\times 10^{10} h^{-1}M_{\odot}$. We can ignore the results below that halo mass scale since the mean number density of halos fall rapidly below that mass scale as seen from figure \ref{fig:HOD_times_hmf_98}.
These considerations show that in the scaling relation model, the ratio $\mHI/m$ is physically reasonable in the vast majority of halos.
\label{lastpage}
\end{document}